\documentclass[journal abbreviation,times,twocolumn,final,authoryear]{elsarticle}

\usepackage{jasr}
\usepackage{framed,multirow}
\usepackage{booktabs}
\usepackage{subfigure}
\usepackage{graphicx}
\usepackage{amssymb}
\usepackage{latexsym}
\usepackage{caption}
\usepackage{makecell}

\usepackage{url}
\usepackage{xcolor}
\definecolor{newcolor}{rgb}{.8,.349,.1}

\usepackage[citebordercolor=white]{hyperref}

\journal{Advances in Space Research}

\begin{document}

\verso{Given-name Surname \textit{etal}}

\begin{frontmatter}

\title{New method for Earth neutral atmospheric density retrieval based on energy spectrum fitting during occultation with LE/\emph{Insight}-HXMT}%

\author[1,2]{Daochun Yu\corref{cor1}}
\ead{yudaochun18@mails.ucas.ac.cn}
\cortext[cor1]{Corresponding author.}
\author[1,2]{Haitao Li\corref{cor1}}
\ead{lihaitao@nssc.ac.cn}
\author[1,2]{Baoquan Li}
\ead{lbq@nssc.ac.cn}
\author[3]{Mingyu Ge}
\ead{gemy@ihep.ac.cn}
\author[3]{Youli Tuo}
\ead{tuoyl@ihep.ac.cn}
\author[3]{Xiaobo Li}
\ead{lixb@ihep.ac.cn}
\author[2,3]{Wangchen Xue}
\ead{xuewc@ihep.ac.cn}
\author[1]{Yaning Liu}
\ead{liuyaning@nssc.ac.cn}

\address[1]{Key Laboratory of Electronics and Information Technology for Space Systerms, National Space Science Center, Chinese Academy of Sciences, Beijing 100190, China}
\address[2]{University of Chinese Academy of Sciences, Beijing 100049, China}
\address[3]{Key Laboratory of Particle Astrophysics, Institute of High Energy Physics, Chinese Academy of Sciences, Beijing 100049,China}

\received{1 May 2013}
\finalform{10 May 2013}
\accepted{13 May 2013}
\availableonline{15 May 2013}
\communicated{S. Sarkar}

\begin{abstract}
We propose a new method for retrieving the atmospheric number density profile in the lower thermosphere, based on the X-ray Earth occultation of the Crab Nebula with the Hard X-ray Modulation Telescope (\emph{Insight}-HXMT) Satellite. The absorption and scattering of X-rays by the atmosphere result in changes in the X-ray energy, and the Earth's neutral atmospheric number density can be directly retrieved by fitting the observed spectrum and spectrum model at different altitude ranges during the occultation process. The pointing observations from LE/\emph{Insight}-HXMT on 16 November 2017 are analyzed to obtain high-level data products such as lightcurve, energy spectrum and detector response matrix. The results show that the retrieved results based on the spectrum fitting in the altitude range of 90--200 km are significantly lower than the atmospheric density obtained by the NRLMSISE-00 model, especially in the altitude range of 110--120 km, where the retrieved results are 34.4\% lower than the model values. The atmospheric density retrieved by the new method is qualitatively consistent with previous independent X-ray occultation results \citep{ref_XRAY_1,ref_XRAY_2}, which are also lower than empirical model predictions. In addition, the accuracy of atmospheric density retrieved results decreases with the increase of altitude in the altitude range of 150--200 km, and the accurate quantitative description will be further analyzed after analyzing a large number of X-ray occultation data in the future.
\end{abstract}


\end{frontmatter}


\section{Introduction}

Accurate measurements of Earth's atmospheric density are very important for maneuver planning, precise orbit determination, satellite lifetime prediction and return control of re-entry vehicle \citep{STORZ20052497,lifttime,book0}. With the increasing demand for the Earth's atmospheric density, various semi-empirical atmosphere models have been developed, such as CIRA model \citep{CIRA}, Jacchia model \citep{1964SAOSR.170.....J,1971SAOSR.332.....J,Jacchia1977ThermosphericTD}, DTM model \citep{BRUINSMA20031053,2012JSWSC...2A..04B,2015JSWSC...5A...1B}, MSIS model \citep{refMSIS86,refMSIS00,refMSIS20}. Due to complex changes in the upper atmosphere, the atmospheric density provided by semi-empirical models tends to have an RMS error of 30\% or more at much higher altitudes relative to the lower thermosphere \citep{error_density,DOORNBOS20081115}, in general, the RMS error near 120 km should be much smaller \citep{EMMERT2015773}.
However, there is very little data on density errors in the lower thermosphere. By analysing NRLMSISE-00 average variation of log density relative to the global mean, \cite{EMMERT2015773} found that in the lower thermosphere near 120 km, the density variation was mainly semidiurnal, and the amplitude of the tidal variations was large. Therefore, it is necessary to develop new measurement methods to accurately measure atmospheric density and calculate density error, such as in-situ measurement method \citep{Eriksen:99,inbook,2016ivs..conf..363T} and remote sensing measurement method \citep{ref_SABER0,Meier2015RemoteSO,ZHANG2020105056}.

Occultation measurement, as a remote sensing detection method, is widely used in the retrieval of atmospheric parameters such as atmospheric density, atmospheric temperature and pressure \citep{KYROLA20041020}. There are stellar occultation methods based on ultraviolet (UV) band \citep{HAYS1973273,1976GeoRL...3..607A,ref_UV_2,ref_UV_3,acp-10-11881-2010}, infrared (IR) band \citep{2009EGUGA..11.2714R,amt-3-523-2010,acp-10-11881-2010}, visible band \citep{acp-10-11881-2010,amt-5-1059-2012} and radio band \citep{leijiuhou,amt-4-1077-2011,chou} to retrieve the Earth's atmospheric density. As a new interdisciplinary technique, X-ray occultation is also used for atmospheric density retrieval. One advantage of density retrieval based on X-ray occultation is that X-ray photons interact directly with electrons in the K- and L-shell of atoms (including atoms in molecules), independent of atmospheric chemistry, atmospheric thermodynamics and ionization, which greatly reduces the mathematical complexity of simulating absorption processes compared with the occultations in the infrared or ultraviolet band \citep{ref_XRAY_1}. In addition, the neutral atmospheric density in the upper mesosphere and lower thermosphere can be obtained by X-ray occultation, which is difficult to detect by other means \citep{ref_SABER0,sparse}.
\cite{ref_XRAY_1} used X-ray occultation sounding for the first time to retrieve atmospheric density, and obtained atmospheric neutral densities in the altitude ranges of 100--120 km and 70--90 km based on PCA/RXTE and USA/ARGOS occultation data, and it was found that the retrieved atmospheric density was smaller than the density from the NRLMSISE-00 model. \cite{ref_XRAY_2} obtained the average neutral atmospheric density at low latitudes in the altitude range of 70--200 km based on 219 occultation profiles with X-ray astronomy satellites Suzaku and Hitomi, and it was found that the retrieved results of atmospheric density in the altitude range of 70--110 km were significantly smaller than the model values of NRLMSISE-00. The difference between retrieved results and the model values may be caused by the long-term accumulation of greenhouse gases, dynamical effects such as gravity waves, the difference of temperature profile \citep{ref_XRAY_1,ref_XRAY_2}. In order to further prove the existence of differences, it is necessary to cross-check by the occultation data of other X-ray satellites or new retrieval methods.

\cite{ref_XRAY_1} obtained the neutral atmospheric density through the lightcurve fitting method. \cite{ref_XRAY_2} obtained the column density first by energy spectrum fitting, and then obtained the atmospheric number density by backward deduction. In this paper, we propose a new spectral fitting method to directly obtain the neutral atmospheric number density in the altitude range of 90--200 km, which is different from the previous method \citep{ref_XRAY_1,ref_XRAY_2}. Therefore, it provides a new means to cross-check the difference between model values and retrieved results. Based on the X-ray occultation observations with LE/\emph{Insight}-HXMT, energy spectrum modeling is performed, and then the Markov chain Monte Carlo (MCMC) method is used to obtain the neutral atmospheric density in the altitude range of 90--200 km.

The paper is structured as follows. The observations and data reduction are described in Sect. \ref{sec:obs_and_dataredunction}. Sect. \ref{sec:Lightcurve_modeling} shows energy spectrum modeling and atmospheric density retrieval. The conclusions are given in Sect. \ref{sec:Conclusions}.

\section{Observation geometry and data reduction} \label{sec:obs_and_dataredunction}
\begin{figure}[t]
	\centering
	\includegraphics[scale=0.3]{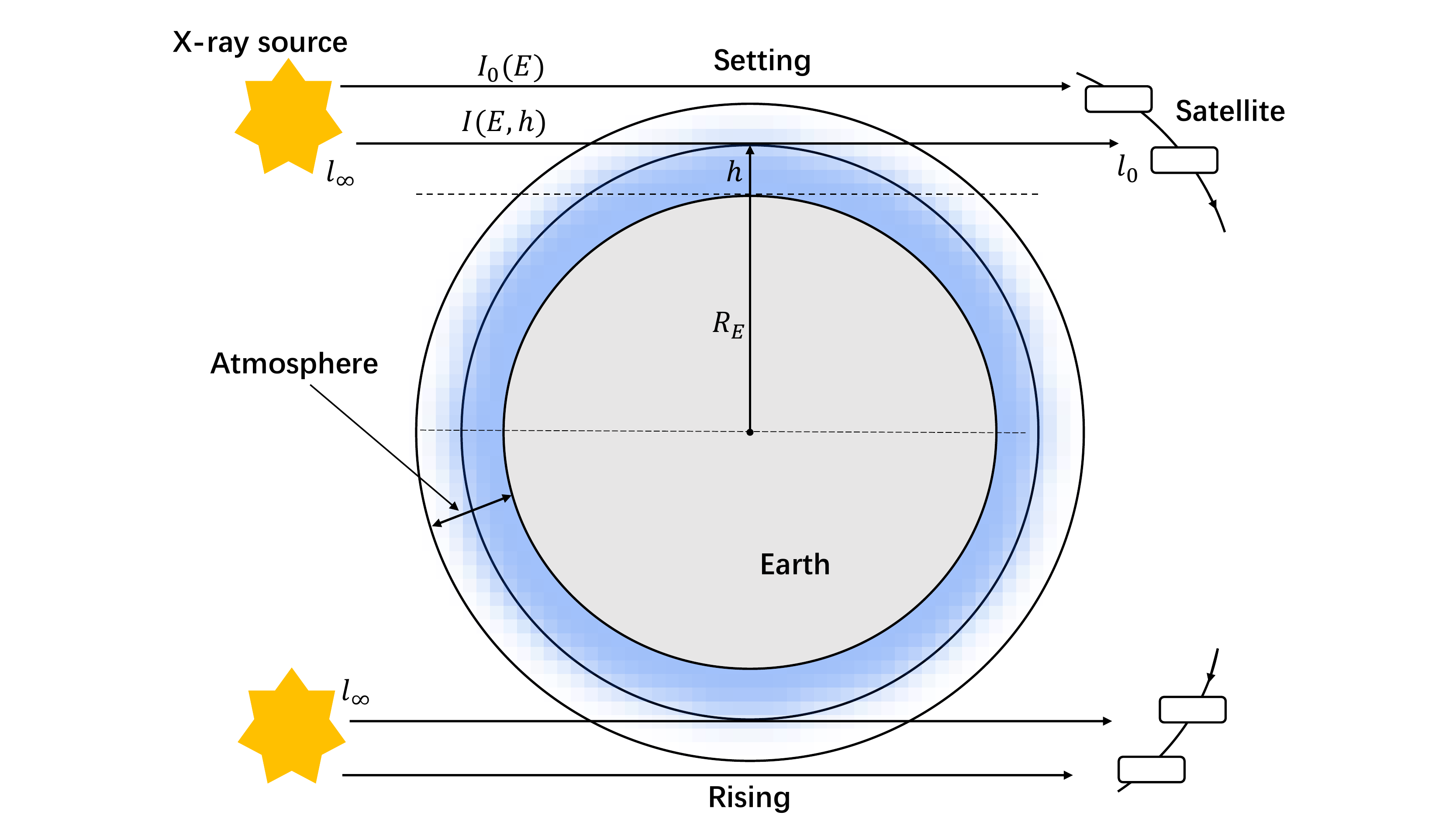}
	\caption{(Color online) Observation geometry of X-ray occultation. $R_E$ is the Earth's radius. Both types of occultation (Rising and Setting) are marked. $I_{0}(E)$ is the unattenuated spectrum, which is not absorbed or scattered by the Earth's atmosphere. $I(E,h)$ is the attenuated spectrum during the occultation process, which is a function of the energy (E) and the tangent point altitude (h). The position of the satellite ($l_0$) and the position of the source ($l_{\infty}$) are also marked.}
	\label{geometry}
\end{figure}

The absorption of X-ray photons increases with increasing atmospheric density, as the tangent point altitude (h) decreases during the occultation process. An X-ray satellite can observe two occultations of the same source per orbital period as the source rises from (or sets behind) the limb of Earth's atmosphere. The observation geometry of X-ray occultation is shown in Figure \ref{geometry}. In this paper, the \emph{Insight}-HXMT measurement of the X-ray occultation of the Crab Nebula is analyzed to retrieve the Earth's atmospheric density.

\emph{Insight}-HXMT is China's first X-ray astronomy satellite, carrying three main scientific payloads, which are the High Energy X-ray telescope (HE), the Medium Energy X-ray telescope (ME) and the Low Energy X-ray telescope (LE) \citep{ref_chinafirst,refHXMT1_1,refHXMT1}. In this paper, because photon extinction in the low-energy X-ray band is obvious, only the observations from the Low Energy X-ray telescope are used for analysis, and as a huge advantage of LE/\emph{Insight}-HXMT, there is no pile-up effect when observing strong sources. The detailed description of the observational data used for analysis in this paper is shown in Table \ref{data_description}. The Crab Nebula is chosen as the target source because of its strong X-ray signal and high stability, and it is often used as a standard candle for calibrating X-ray astronomy satellites \citep{ref_candle,refstandard}. 
\begin{table*}[t]
	\caption{Data description of X-ray occultation of the Crab Nebula.}
	\label{data_description}
	\tabcolsep 5pt
	\begin{tabular}{cccccccccc}
		\toprule[0.1mm]
		Obs ID &{Target}& {Ra}&
		{Dec } & Start time& Stop time & \makecell*[c]{Latitude\\range} & \makecell*[c]{Longitude\\range} & \makecell*[c]{Occultation\\type} \\ 
		& & ($^\circ$)&($^\circ$)&(UTC)&(UTC)& ($^\circ$)&  ($^\circ$) &  & \\\hline
		P0111605008 & Crab Nebula & 83.6330 & 22.0145 & \makecell*[c]{2017-11-16\\T18:54:48} & \makecell*[c]{2017-11-16\\T19:03:05} & 49.07--21.35& -13.44--3.05& Rising  \\\hline
	\end{tabular}
\end{table*}

\begin{figure}[t]
	\centering
	\includegraphics[scale=0.45]{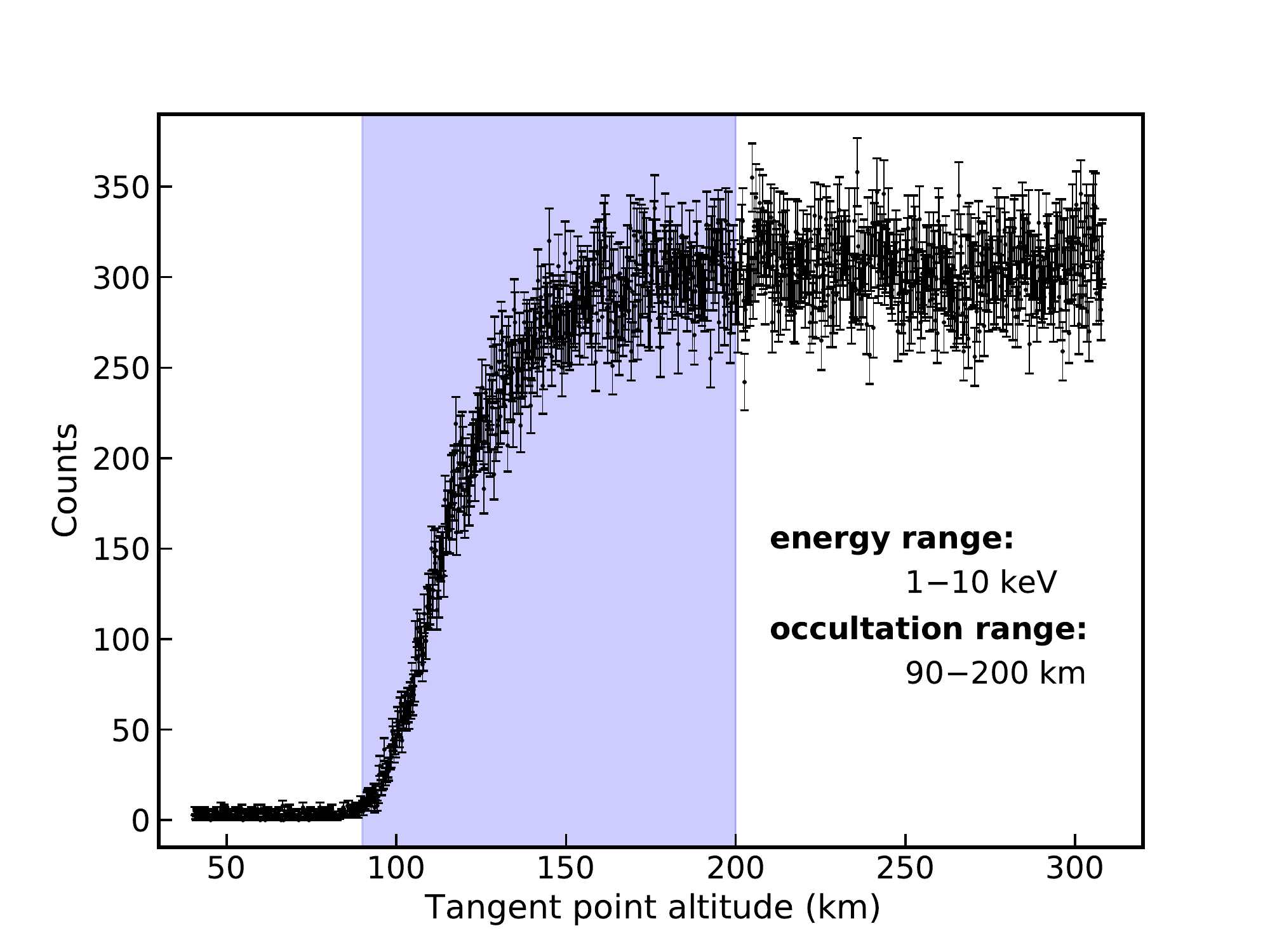}
	\caption{(Color online) The observed lightcurve during the X-ray occultation of the Crab Nebula with LE/\emph{Insight}-HXMT. The lightcurve cover an energy range of 1--10 keV. Black dots with error bars are observed data, and the blue shaded areas mark the occultation range. Extinction begins at 200 km and ends at 90 km.}
	\label{fig:lightcurve}
\end{figure}

\begin{figure}[t]
	\centering
	\includegraphics[scale=0.45]{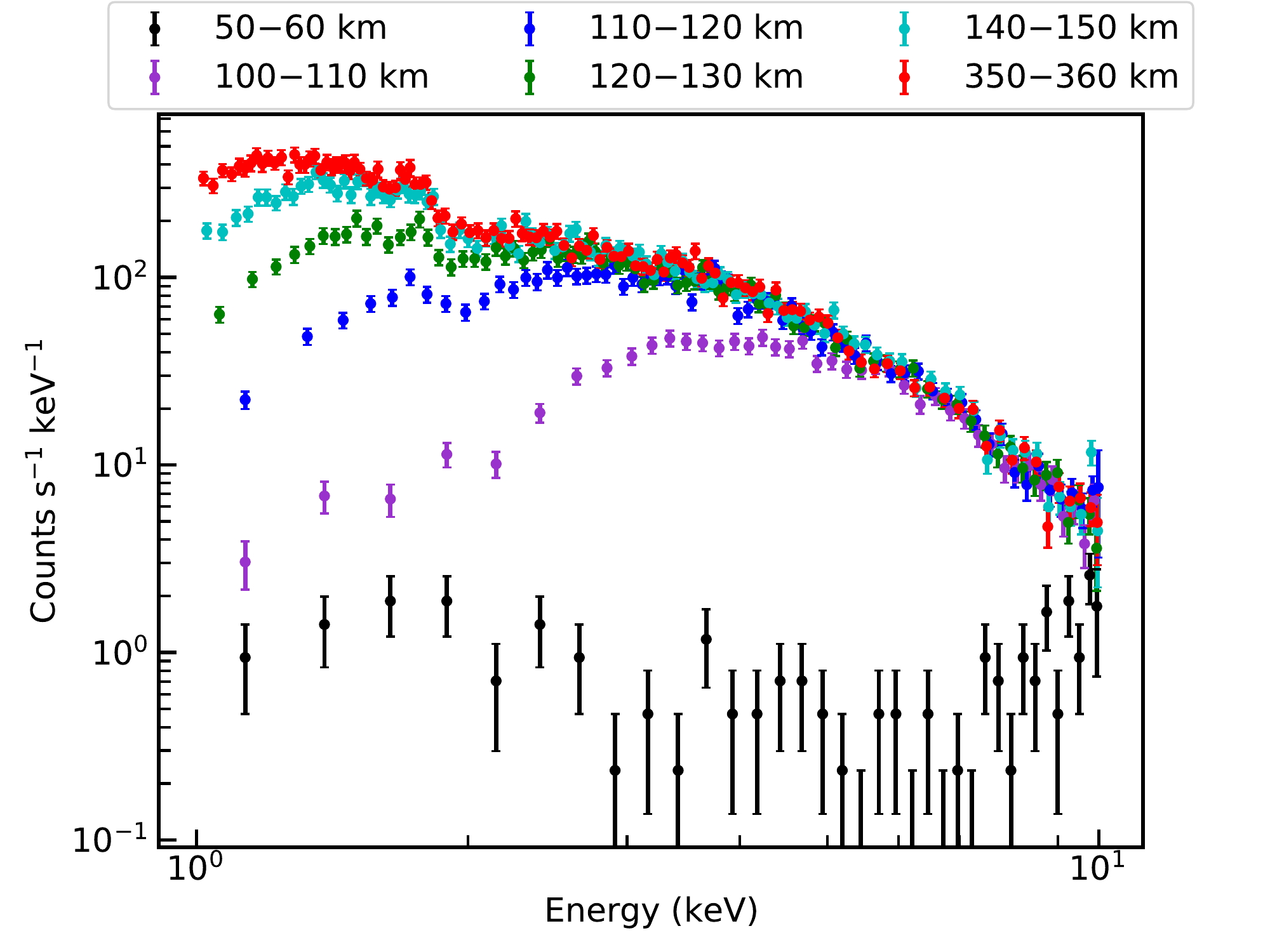}
	\caption{(Color online) Comparison of X-ray energy spectra at different altitude ranges. These energy spectra cover an energy range of 1--10 keV. The red represents the unattenuated X-ray energy spectrum, the black represents the fully attenuated X-ray spectrum, and other colors represent the X-ray energy spectra during the occultation process. However, for clarity, only four X-ray energy spectra are given at different altitude ranges during the occultation process.}
	\label{fig:spectrum}
\end{figure}

Through the data processing software \href{http://hxmtweb.ihep.ac.cn/software.jhtml}{hxmtsoft v2.04} and calibration database \href{http://hxmtweb.ihep.ac.cn/caldb/628.jhtml}{hxmt CALDB v2.05}, the level 1 (1L) pointing observation data of \emph{Insight}-HXMT satellite are processed to obtain high-level data products, such as lightcurve, energy spectrum, detector response matrix and background files. In order to obtain the observed data during the Earth atmosphere occultation of the Crab Nebula, the good time interval (GTI) is screened according to the following criteria: \emph{ELV} (the elevation of a point source above the horizon) less than 10 degrees. Through data reduction, we extract the lightcurve in the energy range of 1--10 keV, as shown in Figure \ref{fig:lightcurve}, where the occultation range is marked by the blue shaded area. It is found that the X-ray photon counts start to decrease at the altitude of 200 km, and the X-ray photon counts are fully attenuated at the altitude of 90 km, so the occultation range of X-ray photons in the energy range of 1--10 keV is 90--200 km. X-ray energy spectrum in the energy range of 1--10 keV is extracted at intervals of 10 km in the occultation range of 90--200 km, as shown in Figure \ref{fig:spectrum}. For clarity, only the observed energy spectra in altitude ranges of 50--60 km, 100--110 km, 110--120 km, 120--130km, 140--150km, and 350--360 km are shown in Figure \ref{fig:spectrum}. Among them, the energy spectrum in the altitude range of 50--60 km is completely attenuated, while the energy spectrum in the altitude range of 350--360 km is unattenuated, and the energy spectrum in other altitude ranges shown in Figure \ref{fig:spectrum} has different degrees of extinction. It is found that the X-ray photon counts decrease gradually with the decrease of the tangent point altitude, especially in the lower energy segment (\textless 5 keV).

\section{Spectrum modeling and density retrieval}\label{sec:Lightcurve_modeling}
The intensity of X-rays decreases with decreasing altitude, which is related to the increase of atmospheric density with the decrease of altitude. The Beer–Lambert law can be used to describe the attenuation of X-ray intensity in the atmosphere,
\begin{equation}
	I(E,h)=I_{0}(E)e^{-\tau(E,h)},
	\label{eq:beer}
\end{equation}
where $I_{0}(E)$ is the unattenuated X-ray spectrum, $I(E,h)$ is the attenuated X-ray energy spectrum by the Earth's atmosphere, which is related to energy ($E$) and tangent point altitude ($h$), $\tau(E,h)$ is optical depth, which is a dimensionless quantity that determines the transmittance level of a material, 
\begin{equation}
	\tau(E,h) = \sum_{k}\int_{l_{0}}^{l_{\infty}}n_{k}(h)\sigma_{k}(E)dl,
	\label{eq:initial}
\end{equation}
where $n_{k}(h)$ is the number density of each atmospheric component ($k$) along the line of sight, in this paper, three atmospheric element components, oxygen (O, O$_2$), nitrogen (N, N$_2$) and argon (Ar), are included, $\sigma_{k}(E)$ is the X-ray cross-section of each atmospheric component, calculated by the XCOM database \citep{XCOM}, $l_{0}$ is the location of the satellite, $l_{\infty}$ is the location of the celestial X-ray source. The detector response matrix \textbf{R} of LE/\emph{Insight}-HXMT is multiplied with the attenuated energy spectrum $I(E,h)$, in order to model the energy spectrum counts on the satellite detector,  
\begin{equation}
	I_{\rm obs}(E,h) = \textbf{R}I(E,h)+B,
	\label{eq:simulation}
\end{equation}
where $B$, as a free parameter, is background noise. In order to retrieve the atmospheric density in a certain altitude range, the optical depth is multiplied by a free parameter $\gamma$, and the value of the free parameter $\gamma$ is obtained through observation and fitting, so as to correct initial value of atmospheric density (retrieved atmospheric density). The transformation of Equation \ref{eq:initial} is as follows, 
\begin{equation}
	\tau(E,h) = \sum_{k}\gamma\int_{l_{0}}^{l_{\infty}}n_{k}(h)\sigma_{k}(E)dl,
	\label{eq:correct}
\end{equation}
where the detailed description of each variable is given in Equation \ref{eq:initial}. In this paper, $\gamma$ is used to correct the density of NRLMSISE-00. 

By combining Equation \ref{eq:simulation} and \ref{eq:correct}, the energy spectrum model in a certain altitude range can be obtained by binning according to the tangent point altitude dimension. A Bayesian method is used to fit the energy spectrum model and observed spectrum to obtain the retrieved results of correction factor $\gamma$, which is based on Bayes' theorem \citep{ref_Bayes}, 
\begin{equation}
	p(\Theta|D,M)=\frac{p(\Theta|M)p(D|\Theta,M)}{p(D|M)},
	\label{eq:Bayes}
\end{equation}
where $\Theta$ is the  parameter vector, in this paper, $\Theta$=\{$\gamma$, $B$\}, $M$ is a model or hypothesis, $D$ is observed data, $p(\Theta|M)$ is the prior, $p(D|\Theta,M)$ is the likelihood, $p(D|M)$ is the Bayesian evidence and $p(\Theta|D,M)$ is the posterior probability distribution of the parameters. Because the X-ray counts satisfy the Poisson distribution, the C statistic is chosen as the Poisson logarithm  likelihood function \citep{ref_statistic1},
\begin{equation}
	C = 2\sum_{i}[D_{i}(\log D_{i}-\log M_{i})+M_{i}-D_{i}],
	\label{eq:C_statistic}
\end{equation}
where $D_{i}$ is the $i$th observation data point, $M_{i}$ is the $i$th model point. Compared with $\chi^2$ statistic, the advantage of C statistic as the log-likelihood function is that the error of the final best-fit parameter is smaller for low-count bins (even one or zero photons) \citep{1989ApJ...342.1207N}.

\begin{figure*}[t]
	\centering
	\includegraphics[scale=0.30]{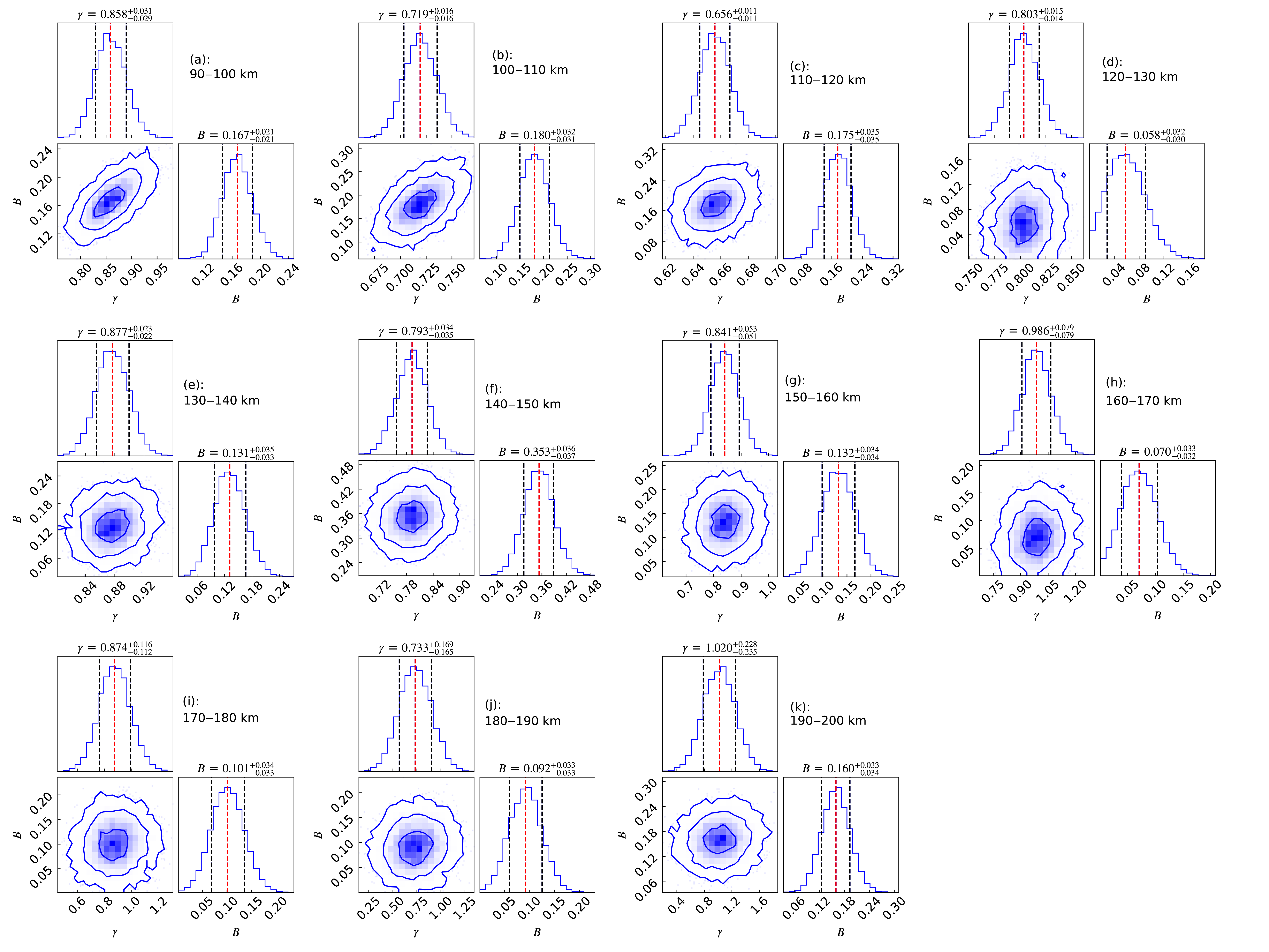}
	\caption{(Color online) Posterior corner plot of the correction factor $\gamma$ and background noise $B$. Panel (a)-(k): the posterior probability distributions of the two free parameters obtained by fitting the observed energy spectrum and the model energy spectrum at different altitude ranges during the occultation process. In each corner plot, the vertical black dashed lines are the quantiles 0.16 and 0.84 of the distribution, the vertical red dashed line is the median of the distribution. The retrieved results of free parameters are shown at the top of the histogram. On the 2-d plots, the confidence intervals for 1-, 2-, 3-$\sigma$ are represented by blue contours.}
	\label{fig:corner}
\end{figure*}

\begin{figure*}[t]
	\centering
	\includegraphics[scale=0.22]{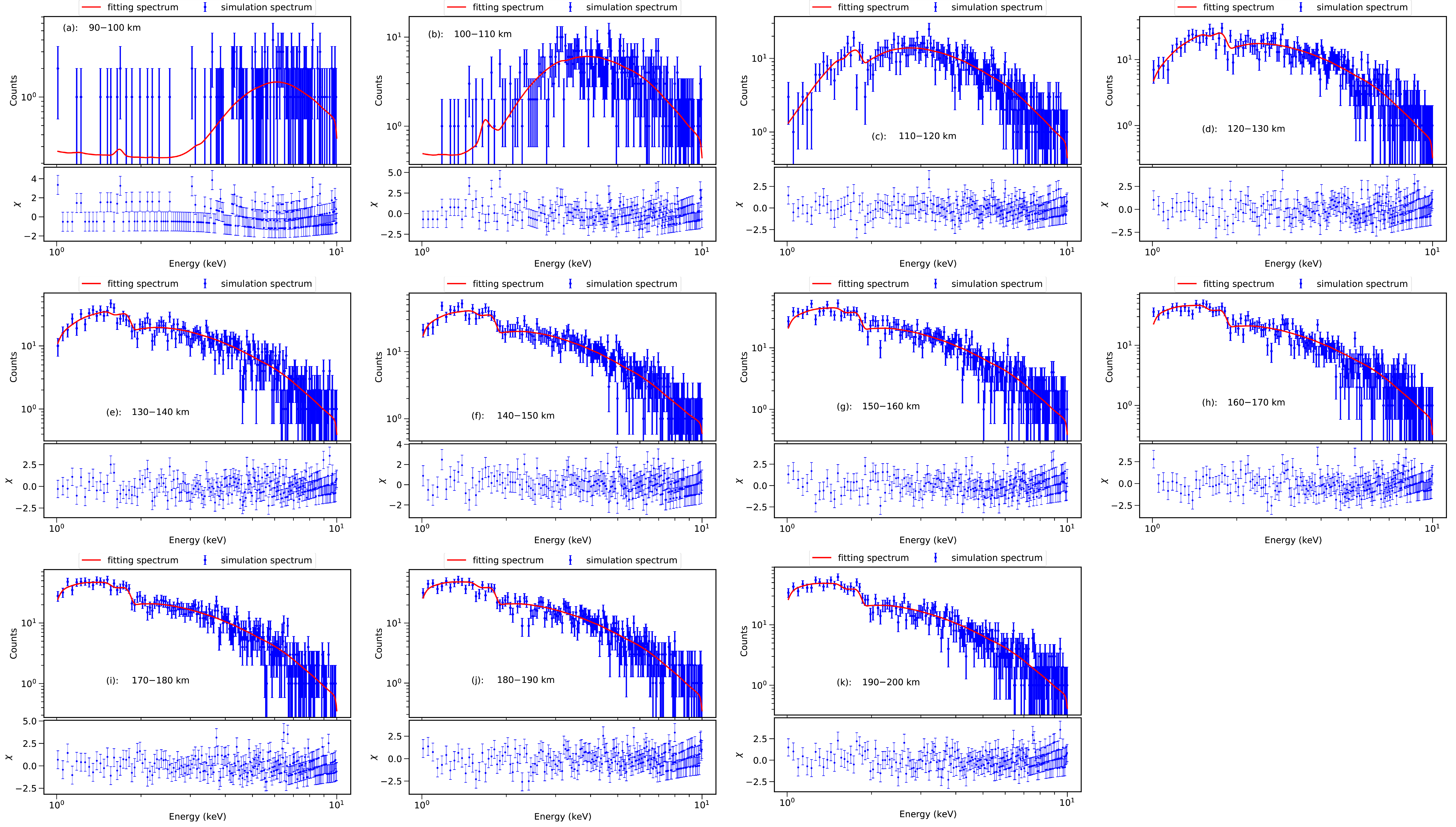}
	\caption{(Color online) Best fit model spectrum and observed spectrum. Panel (a)-(k): the best fit model spectrum and observed spectrum at the different altitude ranges during the occultation process, as well as the residuals between the best fit model and the observed data. In the upper space of each panel, the observed data in blue, and the red solid line represents the best fit model spectrum. And the residuals with error bars of size one between the model and data are shown in the corresponding lower space of each panel.}
	\label{fig:fitting_result}
\end{figure*}

The Markov chain Monte Carlo (MCMC) technique is used to calculate the posterior probability distributions of parameters based on Bayes' theorem. MCMC is a random sampling method in probability density space using Markov chain \citep{ref_MNRAS1,ref_MCMC_annul}. In this paper, we use the {\tt emcee} tool \citep{ref_emcee}, a Python implementation of MCMC, to obtain the posterior probability distributions of parameters based on an affine invariant sampling algorithm \citep{2010CAMCS...5...65G}. A total of 10 walkers are selected in the sampling process, and each walker selects a Markov chain of 10000 steps for sampling. After sampling, the Markov chain for each walker burns the first 1000 steps to obtain the posterior probability distribution of parameters, as shown in Figure \ref{fig:corner}. In each corner plot \citep{ref_corner}, the vertical black dashed lines represent the quantile 0.16 and 0.84 of the distribution, the vertical red dashed line represents the median of the distribution, which is also indicated at the top of the histogram. The best fitting values of the correction factor $\gamma$ and background noise $B$ are also shown in Table \ref{chi2PTE}, where the $\pm$1$\sigma$ of $\gamma$ and $B$ are also given. The best fit model spectrum and observed spectrum at the different altitude ranges during the occultation process are shown in Figure \ref{fig:fitting_result}. In each panel in Figure \ref{fig:fitting_result}, the red solid line represents the best fit model and the blue dots with error bars represent the observed data. The reduced $\chi^2$ \citep{reftestgood} and p-value \citep{2002ApJ...571..545P} are calculated to evaluate the goodness of fit of the best fitting model and observed data, as shown in Table \ref{chi2PTE}. It is found that the best fit is visually in good agreement with the observed data at different altitude ranges during the occultation process.

\begin{table*}[h]
	\caption{The retrieved results of the correction factor $\gamma$ and background niose $B$, and evaluation of goodness of fit between observed energy spectrum and best fit model at different altitude ranges during the occultation process.}
	\label{chi2PTE}
	\tabcolsep 28pt 
	\begin{tabular*}{\textwidth}{ccccccc}
		\toprule[0.1mm] 
		Altitude ranges (km) & $\gamma$ & $B$ (counts)  &{ Reduced $\chi^2$}&{\emph{p}-value} \\ \hline
        \specialrule{0em}{1pt}{1pt}
		90--100        & $0.858^{+0.031}_{-0.029}$          & $0.167^{+0.021}_{-0.021}$                   & 1.008    & 0.423 \\\specialrule{0em}{1pt}{1pt}
		100--110       & $0.719^{+0.016}_{-0.016}$          & $0.181^{+0.032}_{-0.031}$                   & 0.992    & 0.569 \\\specialrule{0em}{1pt}{1pt}
		110--120       & $0.656^{+0.011}_{-0.011}$          & $0.175^{+0.035}_{-0.035}$                   & 0.891    & 0.995 \\\specialrule{0em}{1pt}{1pt}
		120--130       & $0.803^{+0.015}_{-0.014}$          & $0.058^{+0.032}_{-0.030}$                   & 0.987    & 0.617 \\\specialrule{0em}{1pt}{1pt}
		130--140       & $0.877^{+0.023}_{-0.022}$          & $0.131^{+0.035}_{-0.033}$                   & 0.958    & 0.834 \\\specialrule{0em}{1pt}{1pt}
		140--150       & $0.793^{+0.034}_{-0.035}$          & $0.353^{+0.036}_{-0.037}$                   & 1.031    & 0.235 \\\specialrule{0em}{1pt}{1pt}
		150--160       & $0.841^{+0.053}_{-0.051}$          & $0.132^{+0.034}_{-0.034}$                   & 0.941    & 0.915 \\\specialrule{0em}{1pt}{1pt}
		160--170       &
		$0.986^{+0.079}_{-0.079}$          & $0.070^{+0.033}_{-0.032}$                   & 1.000    & 0.495 \\\specialrule{0em}{1pt}{1pt}
		170--180       & $0.874^{+0.116}_{-0.112}$          & $0.101^{+0.034}_{-0.033}$                   & 0.975    & 0.716 \\\specialrule{0em}{1pt}{1pt}
		180--190       & $0.733^{+0.169}_{-0.165}$          & $0.092^{+0.033}_{-0.033}$                   & 1.026    & 0.273 \\\specialrule{0em}{1pt}{1pt}
		190--200       & $1.020^{+0.228}_{-0.235}$          & $0.160^{+0.033}_{-0.034}$                   & 0.997    & 0.519 \\\specialrule{0em}{1pt}{1pt}
		\bottomrule[0.1mm]
	\end{tabular*}
\end{table*}

\begin{figure*}[t]
	\centering
	\includegraphics[scale=0.6]{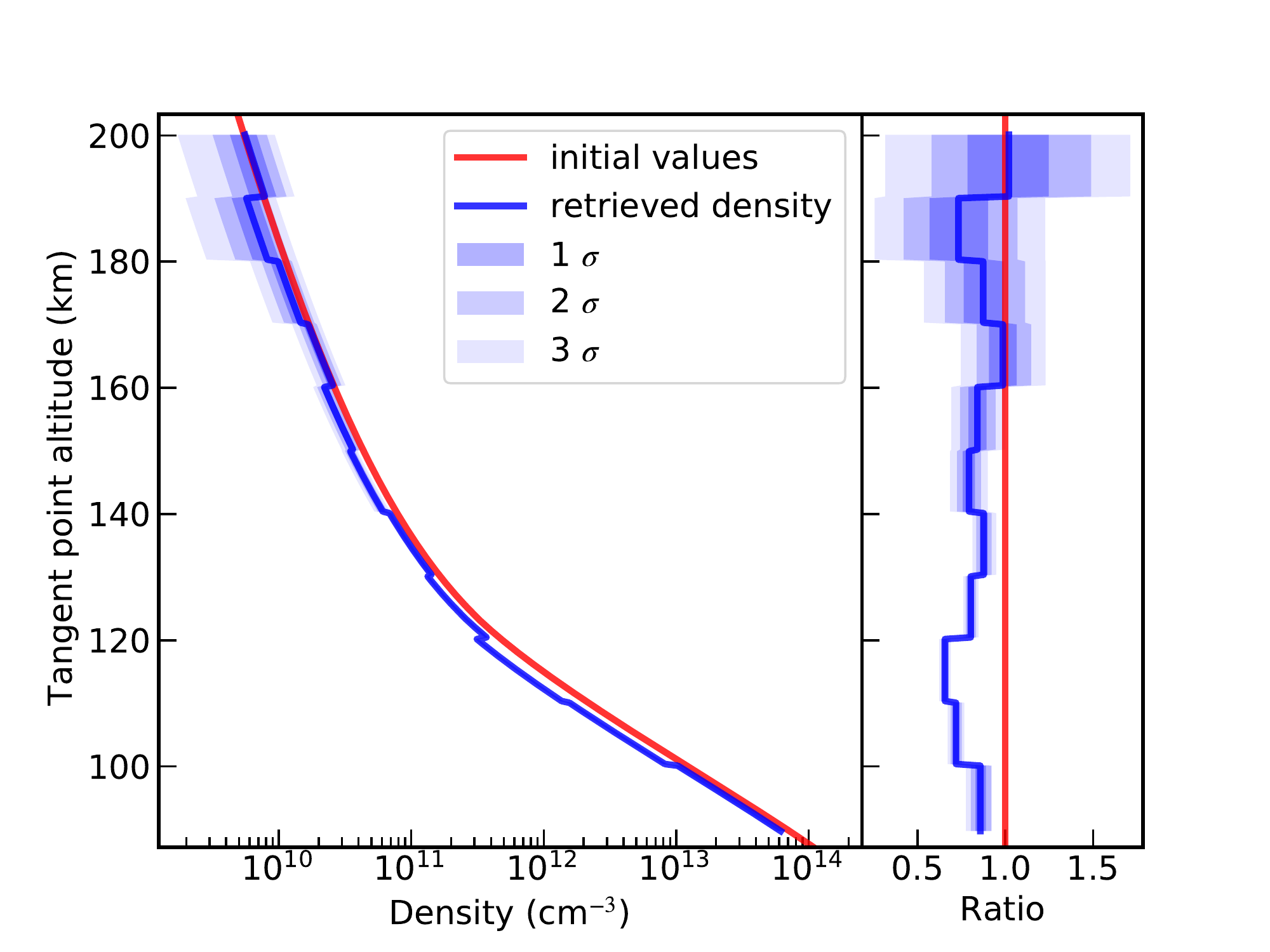}
	\caption{(Color online) Comparison of atmospheric density retrieved results based on energy spectrum fitting and initial values. Left panel: the red solid line represents the initial values of density from the NRLMSISE-00 model, the blue solid line represents the retrieved results that are the product of the correction factor $\gamma$ and the initial values, the 1-, 2-, 3-sigma confidence intervals of the retrieved results are represented by blue shadows of different degrees. Right panel: in order to express the relative relationship between retrieved results and initial values more clearly, both of them are divided by the initial values to achieve normalization.}
	\label{fig:density}
\end{figure*}

Through the above discussion, we obtain the retrieved results of atmospheric density in the altitude range of 90--200 km, as shown in Figure \ref{fig:density}. The red solid line represents the initial values of density, that's the values of $n_{k}(h)$ in Equation \ref{eq:correct}, which are given by the NRLMSISE-00 model. The blue solid line represents the retrieved density, which is the product of the correction factor $\gamma$ and the initial values at different altitude ranges during the occultation process. The 1-, 2-, 3-$\sigma$ confidence intervals of the retrieved results are given by blue shadows from dark colors to light colors. It is found that the retrieved results are generally smaller than the initial values given by the NRLMSISE-00 model, especially in the altitude range of 110--120 km, where the retrieved results are 34.4\% lower than the model values of NRLMSISE-00, so the difference between the initial values of the model and the retrieved density is cross-checked by the new method based on energy spectrum fitting. \cite{refMSIS20} found that NRLMSIS 2.0 N$_{2}$ density in the lower thermosphere is  $\sim$18\% lower than NRLMSISE-00, as a result of colder temperatures in the middle and lower atmosphere. This is consistent with our retrieved results. 
In addition, it is found that the length of confidence interval of retrieved results increases with the increase of tangent point altitude,  especially in the altitude range of 150--200 km. In other words, the accuracy of the retrieved results decreases with  the increase of tangent point altitude in the altitude range of 150--200 km. This is because the extinction is less significant at higher altitude, and most of the photons penetrate the atmosphere, leading to greater uncertainty in the retrieved results.

\section{Conclusions}\label{sec:Conclusions}
In this paper, an energy spectrum modeling method is proposed based on LE/\emph{Insight}-HXMT observations. The atmospheric density retrieved results are obtained by fitting the energy spectrum model and observed data at different altitude ranges during the occultation process based on a Bayesian processing framework, in which the energy spectrum ranges from 1--10 keV. 

It is found that the atmospheric density retrieved results are generally low compared with the a priori values of the NRLMSISE-00 model, especially in the altitude range of 110--120 km, where the retrieved results are 34.4\% lower than the model values of NRLMSISE-00, and the difference between the model values and retrieved results based on the X-ray occultation measurements is further verified by the new method based on energy spectrum fitting. In addition, it is found that the accuracy of atmospheric density retrieved results decreases with the increase of altitude in the range of 150--200 km, which is related to the fact that the extinction is less significant with the increase of altitude. Because atmospheric density decreases with the increase of altitude, the number of X-ray photons absorbed or scattered by atmospheric components also decreases, which brings greater uncertainty to the retrieved results. Therefore, in order to obtain more accurate atmospheric density retrieved results, it is necessary to select an energy band with significant extinction for fitting, and quantitative description of the extinction degree of the energy spectrum during occultation relative to the unattenuated energy spectrum will be analyzed in the future.

Our new retrieval method based on energy spectrum fitting to obtain atmospheric density in the lower thermosphere is different from the previous methods \citep{ref_XRAY_1,ref_XRAY_2}, but it supports previously reported biases between X-ray occultation density and empirical model values. In the future, we can combine this method with previous ones \citep{ref_XRAY_1,ref_XRAY_2} to analyze a large number of occultation data from past, present and future X-ray satellites to obtain the temporal and spatial variation characteristics of the atmospheric density in the lower thermosphere, which can better serve the development of aerospace industry or other scientific fields.

\section{Acknowledgments}
This work was supported by the Youth Innovation Promotion Association CAS (Grant No. 2018178), the National Natural Science Foundation of China (Grant Nos. 41604152, U1938111, U1938109, U1838104, U1838105), the Strategic Priority Research Program on Space Science of Chinese Academy of Sciences (Grant Nos. XDA04060900, XDA15020800, XDA15072103),  the National Key Research and Development Program of China (Grant Nos. 2017YFB0503300, 2016YFA0400800). This work made use of the data from the HXMT mission, a project funded by China National Space Administration (CNSA) and the Chinese Academy of Sciences (CAS).
\bibliographystyle{model5-names}
\biboptions{authoryear}
\bibliography{reference.bib}
\end{document}